\documentclass[prc,superscriptaddress,unsortedaddress,twocolumn,preprintnumbers,amsmath,amssymb,floatfix]{revtex4}
\usepackage{amsmath}
\usepackage{amssymb}
\usepackage{mathrsfs}
\usepackage{bm}
\usepackage{color}
\usepackage{graphicx}
\usepackage{braket}
\usepackage{mathrsfs}
\usepackage{ulem}
\usepackage{here}

\makeatletter
\let\MYcaption\@makecaption
\makeatother
\usepackage{subcaption}
\makeatletter
\let\@makecaption\MYcaption
\makeatother

\makeatletter
\newcommand{\figcaption}[1]{\def\@captype{figure}\caption{#1}}
\newcommand{\tblcaption}[1]{\def\@captype{table}\caption{#1}}
\makeatother

\begin{document}

\title{
Determination of $S_{18}$ from $^{9}$C breakup reaction within a four-body reaction model
}

\author{Shoya Ogawa}
\email[]{ogawa.shoya.615@m.kyushu-u.ac.jp}
\affiliation{Department of Physics, Kyushu University, Fukuoka 819-0395, Japan}

\author{Tokuro Fukui}
\affiliation{Faculty of Arts and Science, Kyushu University, Fukuoka 819-0395, Japan}
\affiliation{RIKEN Nishina Center, Wako 351-0198, Japan}

\author{Jagjit Singh}
\affiliation{Department of Physics and Astronomy, The University of Manchester, Manchester M13 9PL, UK}
\affiliation{Department of Physics, Akal University, Talwandi Sabo, Bathinda, Punjab 151302, India}
\affiliation{Research Center for Nuclear Physics (RCNP), Osaka University, Ibaraki 567-0047, Japan}

\author{Kazuyuki Ogata}
\affiliation{Department of Physics, Kyushu University, Fukuoka 819-0395, Japan}

\date{\today}

\begin{abstract}
 \noindent
 {\bf Background}:
 The astrophysical factor $S_{18}$ for the $^{8}$B($p$,$\gamma$)$^{9}$C has indirectly been measured with the proton removal reactions from $^9$C, 
 elastic breakup of $^9$C off a heavy target, and transfer reactions. 
 Quite recently, the elastic breakup cross section data were reanalyzed with 
 the continuum-discretized coupled channels method (CDCC) assuming a $p+{\rm ^{8}B}$ two-body model for $^9$C and the $S_{18}$ was modified. 
 It was not well justified, however, to treat $^8$B as an inert nucleus given its proton separation energy is only 137~keV.
 \\
 {\bf Purpose}:
 We reexamine the elastic breakup of $^9$C by the four-body CDCC with a $p+p+{\rm ^{7}Be}$ three-body model for $^9$C and evaluate $S_{18}$. 
 To achieve this, we propose a method to disentangle the $p+{\rm ^{8}B}+{\rm ^{208}Pb}$ three-body channel in the four-body CDCC calculation, for the first time.
 \\
 {\bf Methods}:
 We calculate the elastic breakup cross section of $^9$C off a $^{208}$Pb target at 65~MeV/nucleon.
 The obtained breakup cross sections are decomposed into the contributions of 
 the $p+{\rm ^{8}B}+{\rm ^{208}Pb}$ and $p+p+{\rm ^{7}Be}+{\rm ^{208}Pb}$ channels
 by using the solution of the complex-scaled Lippmann--Schwinger equation.
 \\
 {\bf Results}:
 The breakup cross section to the $p+{\rm ^{8}B}+{\rm ^{208}Pb}$ channel reproduces well the shape of the experimental data
 in the low breakup energy region, which is important for determining $S_{18}$.
 By fitting the theoretical result to the experimental data,
 the asymptotic normalization coefficient of $^9$C for the $p+{\rm ^{8}B}$ configuration is determined 
 and we obtain $S_{18}=38.4\pm1.1({\rm theor})\pm5.0({\rm expt})$~eVb.
 \\
 {\bf Conclusion}:
 This result is smaller than the previous value obtained with the three-body CDCC by about 45\%.
 Thus, our new results suggest the necessity of taking into account the fragile nature of $^{8}$B in the $^{9}$C breakup.
\end{abstract}

\maketitle

\section{Introduction}
The proton capture reaction by $^{8}$B, ${\rm ^{8}B}(p,\gamma){\rm ^{9}C}$, 
is important for the explosive hydrogen burning called the hot $pp$ chain~\cite{Wiescher89} in low-metallicity supermassive stars. 
The astrophysical factor $S_{18}$ at zero energy for this reaction has intensively been studied. 
Theoretically, $S_{18}$ has been evaluated with the potential model~\cite{Wiescher89},
the microscopic cluster model~\cite{Descouvemont99}, 
the direct capture model~\cite{Mohr03}, and the Gamow shell model~\cite{Dong23}. 
These studies have yielded significantly varying $S_{18}$ highlighting the sensitivity of the calculations to the chosen theoretical framework and potential parameters. 
Experimentally, due to the difficulty of measuring the reaction at stellar energies, 
several alternative reactions have been utilized to indirectly determine $S_{18}$~\cite{Beaumel01,Trache02,Enders03,Motobayashi03,Guo05,Fukui12,Fukui15}. 
The asymptotic normalization coefficient (ANC) method~\cite{Mukhamedzhanov90} plays a crucial role in determining the $S_{18}$ through indirect measurements.

The elastic breakup reaction using a heavy target nucleus with a strong Coulomb field, such as $^{208}$Pb, 
is one of the alternative reactions that can be used to determine the ANC.
In Ref.~\cite{Ogata06},
it was shown that the breakup cross section obtained within a coupled-channel framework is proportional to the square of the ANC.
The astrophysical factor $S_{17}$ for the ${\rm ^{7}Be}(p,\gamma){\rm ^{8}B}$ reaction was successfully extracted through the analysis of the ${\rm ^{8}B}+{\rm ^{208}Pb}$ reaction; 
the continuum-discretized coupled-channels method (CDCC)~\cite{Yahiro12} was used for calculating the breakup cross section including both nuclear and Coulomb couplings to all orders.

In Ref.~\cite{Fukui12}, this method was applied to the breakup reaction ${\rm ^{208}Pb}({\rm ^{9}C},p{\rm ^{8}B}){\rm ^{208}Pb}$ at 65 MeV/nucleon~\cite{Motobayashi03}.
In the analysis, $^{9}$C was treated as a $p+{\rm ^{8}B}$ two-body system, and the reaction system consists of three particles including the target; we call this reaction model the three-body CDCC.
As a result, $S_{18}=67.3\pm5.4$ eVb was obtained, which is smaller than the value reported in Ref.~\cite{Motobayashi03} by about 13\%.
The authors also reanalyzed the proton removal reaction data~\cite{Trache02} and obtained $S_{18}=63.7\pm13.4$~eVb, 
which modifies the original value of $46\pm6$~eVb reported in Ref.~\cite{Trache02}. 
Thus, a consistent result of $S_{18}$ was obtained from the elastic breakup and proton removal reactions. 
However, it was not clearly justified to treat the weakly bound nucleus $^8$B as an inert core.

Very recently, in Ref.~\cite{Singh21}, the $^{9}$C breakup reaction on a $^{208}$Pb target was investigated with the four-body CDCC, 
which describes $^{9}$C as a $p+p+{\rm ^{7}Be}$ three-body system.
In the four-body CDCC, the breakup states of $^{9}$C are taken into account as pseudostates, 
which are discrete states with positive eigenenergies obtained by diagonalizing a three-body Hamiltonian of $^{9}$C. 
The $^9$C breakup cross sections at 65 and 160~MeV/nucleon were calculated and the resonant and nonresonant contributions to the cross section were discussed. 
However, a comparison with the ${\rm ^{208}Pb}({\rm ^9C},p{\rm ^8B}){\rm ^{208}Pb}$ data has not been made for the following reason.

Because the pseudostates are obtained with a bound-state approximation, 
one cannot differentiate the $p+{\rm ^{8}B}$ and $p+p+{\rm ^{7}Be}$ components. 
In other words, the breakup cross section calculated with the four-body CDCC contains the two following channels:
\begin{align*}
 {\rm ^{9}C}+{\rm ^{208}Pb} &\rightarrow p+{\rm ^{8}B}+{\rm ^{208}Pb}
 ~~
 (\mbox{three-body channel}),
 \\
 {\rm ^{9}C}+{\rm ^{208}Pb} &\rightarrow p+p+{\rm ^{7}Be}+{\rm ^{208}Pb}
 ~~
 (\mbox{four-body channel}).
\end{align*}
Obviously, one must select the former from the four-body CDCC result to make a comparison with 
the ${\rm ^{208}Pb}({\rm ^{9}C},p{\rm ^{8}B}){\rm ^{208}Pb}$ experimental data.

%
This is achieved by employing the solution of the complex-scaled Lippmann-Schwinger equation (CSLS)~\cite{Kikuchi10,Kikuchi11} for decomposing the breakup cross section into the two components.
CSLS has been applied to the analysis of the $d(\alpha,\gamma){\rm ^{6}Li}$ reaction, where the $d+\alpha$ and $p+n+\alpha$ channels exist in the scattering state, 
and it has successfully separated the contributions of these channels in the cross section~\cite{Kikuchi11}.
CSLS was also implemented in the four-body CDCC to investigate the ${\rm ^{6}He}+{\rm ^{12}C}$ reaction, 
for which only the $n+n+\alpha+{\rm ^{12}C}$ four-body breakup channel exists, 
and a dineutron structure in the $2^{+}_{1}$ resonant state was suggested~\cite{Kikuchi13,Ogawa22}.

In this paper, for the first time, we propose a method to calculate breakup cross sections for both the three- and four-body channels by using the four-body CDCC combined with CSLS.
It should be noted that in the four-body breakup channel, 
one needs to describe the continuum state of the three charged particles, $p+p+{\rm ^{7}Be}$, for which a proper asymptotic form is not known. 
Nevertheless, using our method described below, one can calculate the four-body breakup cross section $d\sigma/d\varepsilon$ with $\varepsilon$ being the breakup energy measured from the three-body threshold.
We reanalyze the ${\rm ^{208}Pb}({\rm ^{9}C},p{\rm ^{8}B}){\rm ^{208}Pb}$ experimental data~\cite{Motobayashi03} and determine $S_{18}$.

The construction of this paper is as follows.
In Sec.~II, we describe the theoretical framework. 
In Sec.~III, we present and discuss the numerical results. 
Finally, in Sec.~IV, we give a summary of this study.

\section{Formalism}
\begin{figure}[tbp]
 \centering
 \includegraphics[scale=0.3]{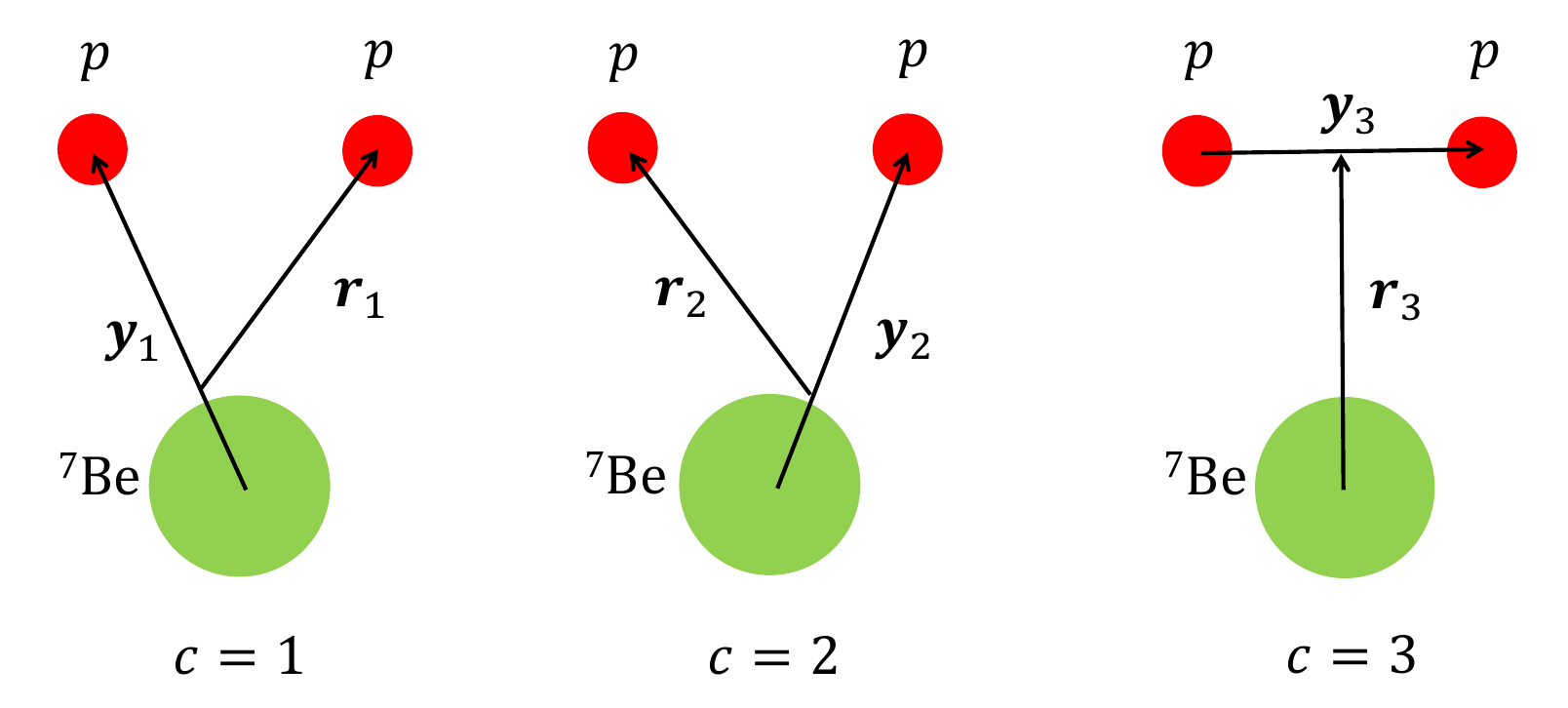}
 \caption{Jacobi coordinates for the three-body system.}
 \label{fig:jacobi}
\end{figure}

We apply the Gaussian expansion method (GEM)~\cite{Hiyama03} to describe the ground state and pseudostates of $^9$C.
In GEM,
the wave function of the three-body system is expanded with the Gaussian bases 
on the Jacobi coordinates as shown in Fig.~\ref{fig:jacobi}.
The bases are given by
\begin{align}
 \psi_{j\lambda}(\bm{y}_c) 
  &=
  y^{\lambda}_{c} e^{-(y_{c}/\bar{y}_{j})^2} Y_{\lambda}(\Omega_{y_c}), 
  \\
 \tilde{\psi}_{i \ell}(\bm{r}_c) 
  &=
  r^{\ell}_{c} e^{-(r_{c}/\bar{r}_{i})^2} Y_{\ell}(\Omega_{r_c}) 
\end{align}
with
\begin{align}
 \bar{y}_{j} &= (\bar{y}_{\rm max}/\bar{y}_{1})^{(j-1)/j_{\rm max}} ,
  \\
 \bar{r}_{i} &= (\bar{r}_{\rm max}/\bar{r}_{1})^{(i-1)/i_{\rm max}} ,
\end{align}
where the index $j$ ($i$) means the $j$th ($i$th) base function for the Jacobi coordinate
$\bm{y}_{c}$ ($\bm{r}_{c}$), the symbol $\lambda$ ($\ell$) denotes the angular momentum 
associated with $\bm{y}_{c}$ ($\bm{r}_{c}$). 
Using the bases, we diagonalize the Hamiltonian:
\begin{align}
 \label{eq:hamiltonian_9C}
 h &= K_{y} + K_{r} 
 + v^{{\rm N+C}}_{p ^{7}{\rm Be}} + v^{{\rm N+C}}_{p ^{7}{\rm Be}} 
 + v^{{\rm N+C}}_{pp} + v_{\rm 3b} + v_{\rm PF}.
\end{align}
Here, $K_y$ ($K_r$) means the kinetic energy operator associated with $\bm{y}$ ($\bm{r}$).
The interactions for the $p$-${\rm ^{7}Be}$ and $p$-$p$ systems are represented as
$v^{{\rm N+C}}_{p ^{7}{\rm Be}}$ and $v^{{\rm N+C}}_{pp}$, respectively.
The superscript N (C) represents the nuclear (Coulomb) part.
The phenomenological three-body force is denoted by $v_{\rm 3b}$.
All interactions in $h$ are the same as those used in Ref.~\cite{Singh21}.
For the valence protons of $^{9}$C, 
the 0$s$ orbit is occupied by the protons in the ${\rm ^{7}Be}$ core. 
This forbidden state $\varphi_{\rm PF}$ can be excluded from the $p+p+{\rm ^{7}Be}$ system by using the so-called pseudo potential~\cite{Saito77} $v_{\rm PF}$
expressed as
\begin{align}
 v_{\rm PF}=\lim_{\lambda_{\rm PF}\rightarrow\infty} \lambda_{\rm PF} \sum_{c=1}^{2}
 \ket{\varphi_{\rm PF}(\bm{y}_{c})} \bra{\varphi_{\rm PF}(\bm{y}_{c})} ,
\end{align}
where $\lambda_{\rm PF}=10^{6}$ MeV is taken in actual calculations. 
The detail is explained in Sec.~3.2 of Ref.~\cite{OgawaPhD}.

The ${\rm ^{9}C}+{\rm ^{208}Pb}$ system is described with a $p+p+{\rm ^{7}Be}+{\rm ^{208}Pb}$ four-body model, 
and the Schr\"{o}dinger equation is
given by
\begin{align}
 \label{eq:Schrodinger-eq}
 \left[ K_{R} + U + h - E \right] \ket{\Psi^{(+)}} = 0 
\end{align}
with
\begin{align}
 U = U^{\rm N}_{p} + V^{\rm C}_{p} + U^{\rm N}_{p} + V^{\rm C}_{p}
 + U^{\rm N}_{^7{\rm Be}} + V^{\rm C}_{^7{\rm Be}} ,
\end{align}
where $\bm{R}$ represents the coordinate between $^{208}$Pb and the center of mass of $^{9}$C.
The operator $K_{R}$ is the kinetic energy associated with $\bm{R}$,
$U^{\rm N}_{p}$ ($U^{\rm N}_{^7{\rm Be}}$) and $V^{\rm C}_{p}$ ($V^{\rm C}_{^7{\rm Be}}$) are the optical potential
and Coulomb interaction between $p$ ($^7$Be) and $^{208}$Pb, respectively.
In the four-body CDCC,
the Schr\"{o}dinger equation in the model space $\mathcal{P}$ is expressed as
\begin{eqnarray}
 \label{eq:CDCC-eq}
 \mathcal{P}\left[ K_{R} + U + h - E \right] \mathcal{P}\ket{\Psi^{(+)}} = 0 
\end{eqnarray}
with
\begin{align}
 \label{eq:model-space}
 \mathcal{P} = \sum_{n} \ket{\Phi_{n}}\bra{\Phi_{n}}.
\end{align}
Here $\Phi_{n}$ is the $n$th pseudostate of $^9$C.
$\mathcal{P}$ is assumed to be a set of the three-body wave functions that is needed to describe the reaction process of our interest; it is sometimes called an approximate complete set. 
In this study, we use the eikonal CDCC~\cite{Ogata06,Ogata03} to solve Eq.~\eqref{eq:CDCC-eq},
where the coupled-channel calculations are performed with the eikonal approximation,
because the standard CDCC calculation with the Coulomb breakup requires high numerical costs.

Solving Eq.~\eqref{eq:CDCC-eq}, we obtain the discretized transition matrix represented as
\begin{align}
 \label{eq:tmat-dis}
 T_{n} = \braket{\Phi_{n}\chi^{(-)}_{n}(\bm{P}_{n})|U|\mathcal{P}\Psi^{(+)}} ,
\end{align}
where $\chi^{(-)}_{n}(\bm{P}_{n})$ is the incoming Coulomb wave function with the asymptotic momentum $\bm{P}_{n}$.
To obtain the breakup cross section for the $p+{\rm ^{8}B}+{\rm ^{208}Pb}$ three-body channel, 
we calculate the following continuous transition matrix:
\begin{align}
 \label{eq:tmat-conti}
 T_{\varepsilon}(\bm{k},\bm{P})  = \sum_{n} f_{n,\varepsilon}(\bm{k}) T_{n} 
\end{align}
with the smoothing factor defined as
\begin{align}
 \label{eq:smoothing}
 f_{n,\varepsilon}(\bm{k})
 =
 \braket{\Phi^{(-)}_{\varepsilon}(\bm{k})|\Phi_{n}} .
\end{align}
Here $\Phi^{(-)}_{\varepsilon}$ is the incoming three-body scattering wave with continuum energy $\varepsilon$,
for which the incident channel corresponds to the $p+{\rm ^{8}B}$ with the relative momentum $\bm{k}$ between $p$ and ${\rm ^{8} B}$,
and $\bm{P}$ is the asymptotic momentum regarding $\bm{R}$.
In CSLS, the smoothing factor is expressed as
\begin{align}
 \label{eq:CSLS-smoothing}
 f_{n,\varepsilon}(\bm{k})
 &=
 \braket{\varphi_{\rm ^{8}B}\phi^{\rm C(-)}_{\bm{k}}|\Phi_{n}} 
 \nonumber \\
 &~~~~
 +
 \sum_{\nu}
 \braket{\varphi_{\rm ^{8}B}\phi^{\rm C(-)}_{\bm{k}}|VU^{-1}_{\theta}|\Phi^{\theta}_{\nu}}
 \frac{1}{\varepsilon-\varepsilon^{\theta}_{\nu}}
 \braket{\tilde{\Phi}^{\theta}_{\nu}|U_{\theta}|\Phi_{n}} 
\end{align}
with
\begin{align}
 V = v^{{\rm N+C}}_{p ^{7}{\rm Be}} + v^{{\rm N+C}}_{pp} + v_{\rm 3b} - v^{\rm C}_{p ^{8}{\rm B}},
\end{align}
where $v^{\rm C}_{p ^{8}{\rm B}}$ is the Coulomb interaction between $p$ and $^{8}$B,
$\varphi_{\rm ^{8}B}$ is the ground state wave function of ${\rm ^{8}B}$,
$\phi^{\rm C(-)}_{\bm{k}}$ is the incoming Coulomb wave function of the $p$-${\rm ^{8}B}$ system.
The complex scaling operator $U_{\theta}$ is defined by $\bm{y}\rightarrow\bm{y}e^{i\theta}$ and $\bm{r}\rightarrow\bm{r}e^{i\theta}$
with the scaling angle $\theta$~\cite{Aguilar71,Balslev71}.
Diagonalizing the scaled Hamiltonian $h^{\theta}\equiv U_{\theta}hU^{-1}_{\theta}$ by using GEM,
we obtain the scaled discretized state $\Phi^{\theta}_{\nu}$ with the complex energy $\varepsilon^{\theta}_{\nu}$ 
and its biorthogonal state $\tilde{\Phi}^{\theta}_{\nu}$.
For efficient numerical calculation, $\phi^{\rm C(-)}_{\bm{k}}$ is described as
\begin{align}
 \bra{\phi^{\rm C(-)}_{\bm{k}}} = \sum_{i} \braket{\phi^{\rm C(-)}_{\bm{k}}|\phi_{i}} \bra{\phi_{i}} .
\end{align}
Here $\phi_{i}$ is the discretized Coulomb wave function. 
This state is obtained by diagonalizing the two-body Hamiltoanian $h_{p{\rm ^8B}}$ of the $p$-$^8$B system, 
which is given by
\begin{align}
 h_{p{\rm ^8B}} = K_{p ^{8}{\rm B}} + v^{\rm C}_{p ^{8}{\rm B}},
\end{align}
with GEM. 
The kinetic energy operator of the system is represented as $K_{p ^{8}{\rm B}}$.
Using Eq.~\eqref{eq:tmat-conti}, the breakup cross section for the three-body channel
is given by
\begin{align}
 \label{eq:dbu-p8B}
 \frac{d\sigma_{p{\rm ^8 B}}}{d\varepsilon}
 &=
 \int d\bm{k} d\bm{P} |T_{\varepsilon}(\bm{k},\bm{P})|^2
 \nonumber \\
 &~~~~\times
 \delta\left(E_{\rm tot} - \frac{\bm{P}^2}{2\mu} - \varepsilon \right)
 \delta\left(\varepsilon - \frac{\bm{k}^2}{2\mu_{p\text{-}{\rm ^{8} B}}} - \varepsilon_{\rm ^{8}B} \right),
\end{align}
where $E_{\rm tot}$ is the total energy of the reaction system,
$\varepsilon_{\rm ^{8}B}$ is the binding energy of $^{8}$B,
$\mu$ and $\mu_{p\text{-}{\rm ^{8}B}}$ are the reduced masses of the $^{9}$C-$^{208}$Pb and 
$p$-$^{8}$B systems, respectively.

As mentioned above, 
it is not possible to calculate directly the smoothing factor for the four-body channel.
Fortunately, in the same way as in the previous studies~\cite{Matsumoto10,Singh21},
the sum of the cross sections corresponding to the three-body and four-body channels is obtained by
\begin{align}
 \label{eq:dbu}
 \frac{d \sigma}{d\varepsilon}
 =
 \frac{1}{\pi} {\rm Im} \sum_{\nu} 
 \frac{T^{\theta}_{\nu} \tilde{T}^{\theta}_{\nu}}{\varepsilon-\varepsilon^{\theta}_{\nu}}
\end{align}
with
\begin{align}
 \tilde{T}^{\theta}_{\nu}
 &=
 \sum_{n} \braket{\tilde{\Phi}^{\theta}_{\nu}|U_{\theta}\Phi_{n}}T_{n},
 \\
 T^{\theta}_{\nu}
 &=
 \sum_{n} T^{*}_{n}\braket{\Phi_{n}|U^{-1}_{\theta}|\Phi^{\theta}_{\nu}}.
\end{align}
Thus, one can calculate the breakup cross section for the four-body channel as
\begin{align}
 \label{eq:dbu-pp7Be}
 \frac{d\sigma_{pp{\rm ^7 Be}}}{d\varepsilon} 
 =
 \frac{d\sigma}{d\varepsilon} - \frac{d\sigma_{p{\rm ^8 B}}}{d\varepsilon} .
\end{align}
%

\section{Results and Discussion}

\subsection{Analysis of $^9$C breakup reaction}
In this study, the optical potentials in Eq.~\eqref{eq:Schrodinger-eq} are constructed by a microscopic folding model 
and can be found in the addendum provided as supplemental material~\cite{Ogawa25_add}.
The Melbourne nucleon-nucleon $g$ matrix~\cite{Amos00} and the Hartree-Fock one-body densities of $^{7}$Be and $^{208}$Pb calculated with the Gogny D1S force~\cite{Minomo10} are adopted. 
The $0^+$, $1^{-}$, and $2^+$ states of $^{9}$C below $\varepsilon=7$ MeV are included in numerical calculations.
For the parameters of Gaussian bases of $\Phi_{n}$,
we use set I in Table I of Ref.~\cite{Singh21}, and set II for $\Phi^{\theta}_{\nu}$.
The scaling angle $\theta$ is set to $17.5^{\circ}$ in Eq.~\eqref{eq:CSLS-smoothing},
and $20^{\circ}$ in Eq.~\eqref{eq:dbu}.
We have confirmed the convergence of numerical results with these model space parameters.

\begin{figure}[tbp]
 \centering
 \includegraphics[scale=0.7]{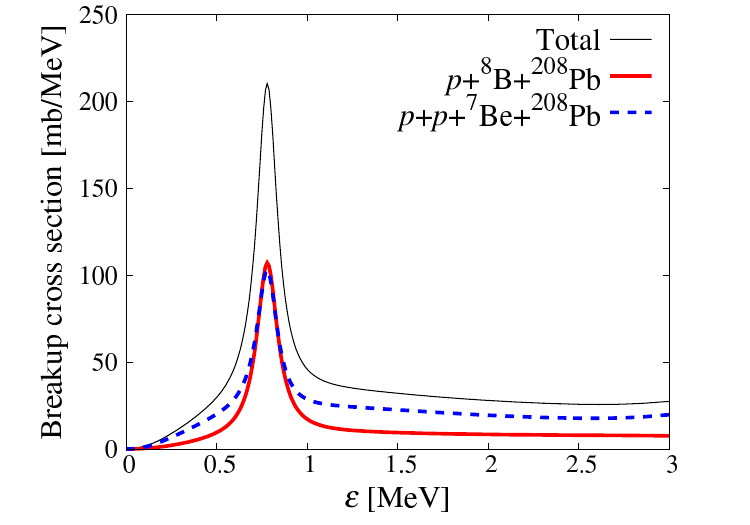}
 \caption{Breakup cross sections of the ${\rm ^{9}C}+{\rm ^{208}Pb}$ reaction at 65 MeV/nucleon.
 The solid and dashed lines describe the breakup cross sections to the three-body and four-body channels, respectively.
 The thin solid line is the sum of them.}
 \label{fig:dbu}
\end{figure}

We show the breakup cross section of the ${\rm ^{9}C}+{\rm ^{208}Pb}$ reaction at 65 MeV/nucleon in Fig.~\ref{fig:dbu}.
The solid and dashed lines represent the breakup cross sections to the three-body and four-body channels, respectively.
The thin solid line is the sum of them.
It is found that the contributions of the three- and four-body channels are comparable near the peak, which comes from the $2^{+}_{1}$ resonant state~\cite{Singh21}. 
This seems to suggest that the $2^{+}_{1}$ state contains the $p+{\rm ^{8}B}$ and $p+p+{\rm ^{7}Be}$ components with almost equal probabilities.
However, it will be too early to draw this conclusion because the role of the final state interaction has not been clarified.
A detailed analysis is ongoing and will be reported in a forthcoming paper.

\begin{figure}[tbp]
 \centering
 \includegraphics[scale=0.7]{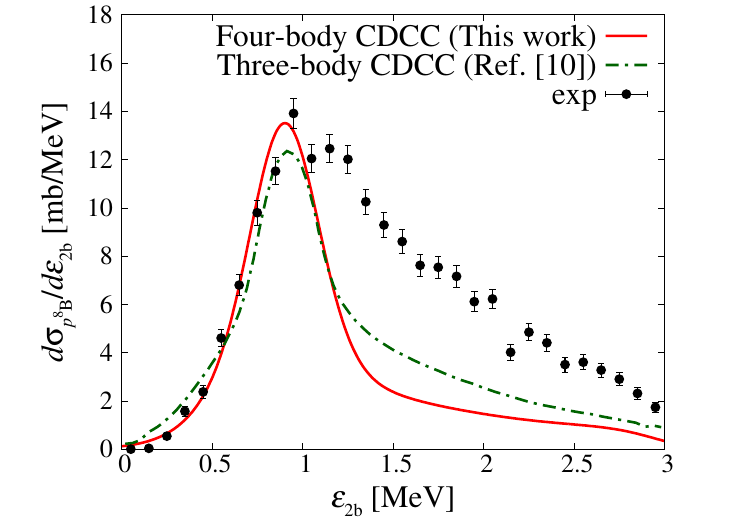}
 \caption{Breakup cross section of $^{208}$Pb($^{9}$C, $p{\rm ^{8}B}$)$^{208}$Pb
 at 65 MeV/nucleon. The result with the three-body CDCC with a normalization factor 1.10
 and the experimental data are taken from Refs.~\cite{Fukui12} and \cite{Motobayashi03},
 respectively.}
 \label{fig:dbu-eff}
\end{figure}
\begin{figure}[tbp]
 \centering
 \includegraphics[scale=0.5]{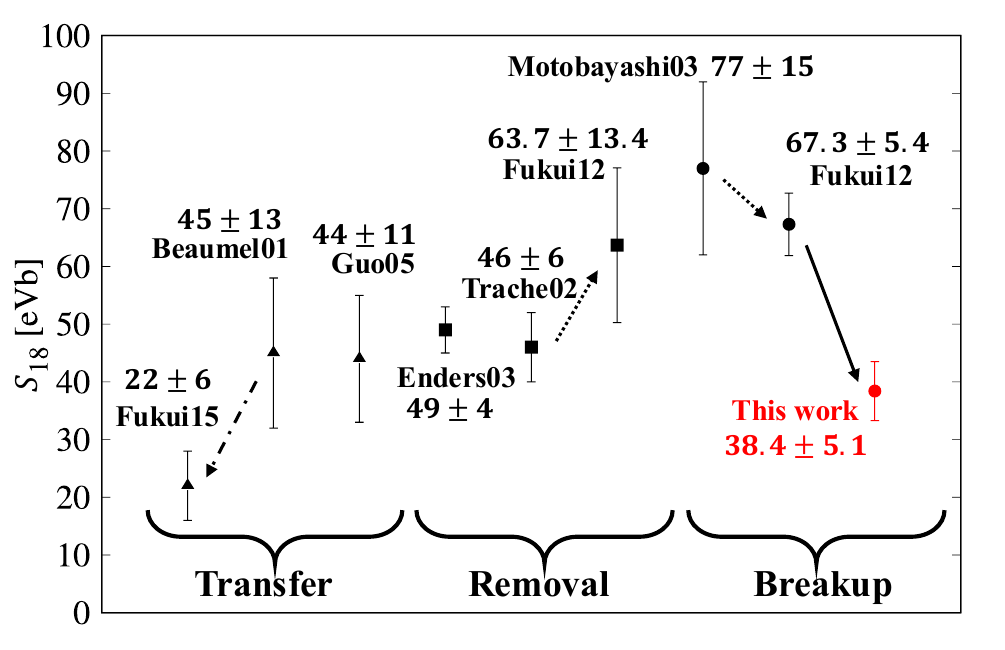}
 \caption{A summary of $S_{18}$ obtained from indirect measurements.
 The previous results represented by triangles, squares, and circles are taken from Refs.~\cite{Fukui15,Beaumel01,Guo05}, \cite{Enders03,Trache02,Fukui12}, 
 and \cite{Motobayashi03,Fukui12}, respectively. See the text for details.}
 \label{fig:Sfac}
\end{figure}
Figure~\ref{fig:dbu-eff} shows the breakup cross sections of the ${\rm ^{208}Pb}({\rm ^{9}C},p{\rm ^{8}B}){\rm ^{208}Pb}$ reaction as functions of ${\varepsilon_{\mathrm{2b}}\equiv \varepsilon - \varepsilon_{^{8}\mathrm{B}}}$.
The solid line represents the result with Eq.~\eqref{eq:dbu-p8B} including the experimental efficiency
and resolution in the same manner as that in Ref.~\cite{Fukui12}.
To determine the ANC, this cross section has been multiplied by a normalization factor 0.797; the experimental data~\cite{Motobayashi03} for $\varepsilon_{\rm 2b}\le 1$ MeV,
which is important for determining $S_{18}$ is used in the fitting.
The dot-dashed line is the result obtained with the three-body CDCC multiplied by 1.10, 
which is taken from Ref.~\cite{Fukui12}.
The four-body CDCC result is in better agreement with the experimental data for $\varepsilon_{\rm 2b}\le 1$ MeV than the three-body CDCC.
At higher $\varepsilon_{\rm 2b}$, the breakup cross section obtained with the four-body CDCC is smaller than that with the three-body CDCC.
This is due to the loss of the flux to the four-body breakup channel, in which $^8$B breaks up into $p+{\rm ^{7}Be}$.
The same trend was seen in the total breakup cross sections of ${\rm ^{3}He}$,
which includes both the $d+p+{\rm target}$ and $p+p+n+{\rm target}$ channels~\cite{Ogawa23}.
Both CDCC results underestimate the experimental data in the high $\varepsilon_{\rm 2b}$ region.
This behavior is also found in the analysis of the $^{6}$He breakup reaction~\cite{Matsumoto10}.
A possible reason for this undershooting is that the target excitation in the final channel is not taken into account in the CDCC calculations.

\subsection{Extraction of ANC from $^9$C breakup reaction}
The single particle ANC $C^{\rm {^{9}C}(s.p.)}_{p{\rm ^8B}}$ is given by
\begin{align}
 \braket{\varphi_{\rm ^{8}B}|\Phi_{\rm ^{9}C}}_{\bm{y}_{1}}
 \sim
 C^{\rm {^{9}C}(s.p.)}_{p{\rm ^8B}}
 \frac{W_{-\eta,\ell_{0}+1/2}(2k_{0}r_{1})}{r_{1}},
 ~~
 r_{1} > r_{N},
\end{align}
where $r_{N}$ is the range of the nuclear interaction between $p$ and $^8$B,
$\Phi_{\rm ^{9}C}$ is the ground state wave function of $^{9}$C,
$k_{0}$ is the relative momentum between $p$ and $^8$B in the ground state of $^9$C,
$W_{-\eta,\ell_{0}+1/2}$ is the Whittaker function, 
and $\ell_{0}$ is the orbital angular momentum of the $p$-$^8$B relative motion in the ground state of $^9$C; 
$\ell_{0}=1$ in the present case.
The subscript $\bm{y}_{1}$ of $\braket{\cdots}$ means the integral variable.
It should be noted here that in our three-body calculation, $^7$Be is treated as inert and its spin is neglected. 
To obtain a realistic ANC,
$C^{\rm {^{9}C}(s.p.)}_{p{\rm ^8B}}$ should be multiplied by the square root of the normalization factor of the cross section.
For details, see Ref.~\cite{Ogata06}.

From the abovementioned normalization factor 0.797 for the four-body CDCC result, 
the ANC $C^{^{9}{\rm C}}_{p{\rm ^8B}}$ is determined to be 0.991~${\rm fm}^{-1/2}$, 
which quite agrees well with the recent result 1.125~${\rm fm}^{-1/2}$ of the variational Monte Carlo calculation~\cite{Nollett11}. 
We then change the geometric parameters of $v_{p{\rm ^{7}Be}}^{\rm N}$ for the ground state by 20\% as in Ref.~\cite{Fukui12} to estimate the fluctuation of the result
due to the violation of the peripherality. 
This turns out to lead to a theoretical uncertainty of approximately 3\% in the $S_{18}$ value.
In addition, taking into account the 13\% systematic uncertainty of the experimental data~\cite{Motobayashi25}, 
the final result for $S_{18}$ becomes $38.4\pm1.1({\rm theor})\pm5.0({\rm expt})$ eVb.
Figure~\ref{fig:Sfac} is a summary of $S_{18}$ obtained from indirect measurements so far.
As mentioned in the introduction part, Fukui and collaborators~\cite{Fukui12} modified (shown with dotted arrow lines in Fig.~\ref{fig:Sfac}) the result of Ref.~\cite{Motobayashi03} to $S_{18}=67.3\pm5.4$~eVb and that of Ref.~\cite{Trache02} to $63.7\pm 13.4$~eVb, 
showing a good consistency between the two values extracted from the elastic breakup and proton removal reactions. 
The present result of $38.4\pm1.1({\rm theor})\pm5.0({\rm expt})$~eVb can be understood as a remodified value of the elastic breakup result of $^9$C by taking into account the fragileness of $^8$B {\lq\lq}core'' in $^9$C (shown with solid arrow line in Fig.~\ref{fig:Sfac}). 
In Fig.~\ref{fig:Sfac}, the result of Ref.~\cite{Enders03} determined from another set of proton removal reaction data is also shown. 
Analysis of the proton removal reaction data~\cite{Trache02,Enders03} with a four-body reaction model will be interesting and important. 
Finally, we remark the results deduced from the nucleon transfer reaction. 
In Refs.~\cite{Beaumel01} and \cite{Guo05}, ${\rm ^{8}B}(d,n){\rm ^{9}C}$ and ${\rm ^{8}Li}(d,p){\rm ^{9}Li}$ were used to extract the ANC $C^{{\rm ^{9}C}}_{p{\rm ^{8}B}}$;
in the latter the charge symmetry for the ANC was assumed. 
These results show a good agreement. 
Later, Fukui and collaborators~\cite{Fukui15} showed that the transfer process through breakup states of $d$ in the initial channel or those of $^9$C in the final channel enhances the ${\rm ^{8}B}(d,n){\rm ^{9}C}$ cross section and reduces $S_{18}$ (shown with dot-dashed arrow line in Fig.~\ref{fig:Sfac}).
Again, a reanalysis of the transfer cross section with a three-body structure model of $^9$C will be an important future work.

\section{Summary}
We have disentangled for the first time the reaction channels in the breakup of the three-body projectile $^9$C, 
which has a bound state in the subsystem, within the four-body reaction framework. 
We employed the four-body CDCC and CSLS and calculated the breakup cross sections of $^9$C off $^{208}$Pb at 65~MeV/nucleon to the $p+{\rm ^{8}B}+{\rm {}^{208}Pb}$ and $p+p+{\rm ^{7}Be}+{\rm ^{208}Pb}$ channels. 
The former was compared with the experimental data and the ANC $C^{^9{\rm C}}_{p^8{\rm B}}$ was determined, 
which resulted in the astrophysical factor of ${\rm ^{8}B}(p,\gamma){\rm ^{9}C}$ at zero energy: $S_{18}=38.4\pm1.1({\rm theor})\pm5.0({\rm expt})$~eVb. 
This result is significantly lower than the previous value obtained with the three-body CDCC assuming the $p+{\rm ^{8}B}$ two-body structure of $^9$C. 
This indicates that the breakup of $^8$B must be taken into account in the description of the $^9$C breakup reactions. 
To draw a firm conclusion on $S_{18}$, reanalyses of the proton removal and nucleon transfer reactions with the $p+p+{\rm ^{7}Be}$ three-body structure model for $^9$C will be necessary.

As mentioned around Eq.~\eqref{eq:model-space}, the eigenstates with positive energies calculated with GEM can be regarded as continuum states of few-body systems in finite space being relevant to the reaction processes. 
In order to calculate physics observables, however, specification of a proper boundary condition in the description of multi-channel reactions is crucial, which has been achieved by the present framework combining CDCC and CSLS. 
This can be extended to more than four-body breakup reactions, 
once a set of eigenstates of many-nucleon systems is provided with GEM. 
Moreover, if we do not distinguish the target particle (especially proton) from the constituents of the projectile, 
all reaction channels including rearrangement channels can be treated on an equal footing.  
The work in this direction is ongoing and will be reported elsewhere.

\section*{Acknowledgments} 

The authors would like to thank T.~Matsumoto for helpful discussions.
S.O. and K.O. thank S.~Watanabe and Y.~Chazono for useful comments.
S.O. acknowledges S.~Ishikawa for valuable comments and discussions.
This work is supported in part by Grant-in-Aid for Scientific Research
(No.\ JP21H04975 [S.O. and K.O.], No.\ JP21K13919 [T.F.], and No.\ JP23KK0250 [T.F.]) from Japan Society for the Promotion of Science (JSPS), 
and JST ERATO Grant No. JPMJER2304, Japan [T.F. and K.O.],
and by the U.K. Science and Technology Funding Council (Grant No. ST/V001116/1 and No. ST/Y000323/1) [J.S.].
J.S. and K.O. acknowledge the support from the University of Manchester-Osaka University 2024 seed-corn fund.  

\bibliography{./ref}

\end{document}